\documentclass[twoside]{photon2007}
\usepackage[latin1]{inputenc}
\usepackage[dvips]{graphicx,epsfig,color}
\usepackage{wrapfig,rotating}
\usepackage{amssymb,amsmath,array}

\begin{document}
\title{
Discussion on the recent proton-DVCS results of Jefferson Lab.} 
\author{Michel Guidal
\vspace{.3cm}\\
Institut de Physique Nucl\'eaire d'Orsay\\
91405 Orsay, FRANCE 
}

\maketitle

\begin{abstract}
We present the recent data issued from the Halls A and B of Jefferson Laboratory
for the Deep Virtual Compton Scattering Process on the proton. An important
set of data for beam spin asymmetries,
unpolarized cross sections and differences of polarized cross sections have been obtained.
We modestly attempt a first ``global" analysis of these three observables at a single 
$<x_B>$, $<Q^2>$ and $<-t>$ kinematic point. We find that it is extremely challenging to 
describe simultaneously these data in the framework of a few Generalized Parton Distributions models.
\end{abstract}

\section{Introduction}

Generalized Parton Distributions (GPDs) have emerged this past decade as a
powerful concept and tool to study nucleon structure. They describe,
among other aspects, the (correlated) spatial and momentum distributions 
of the quarks in the nucleon (including the polarization aspects), its
quark-antiquark content, a way to access the orbital momentum of the quarks, 
etc... We refer the reader to Refs~\cite{goeke,revdiehl,revrady} for example, 
for very detailed and quasi-exhaustive reviews on the GPD formalism and the 
definitions of some of the variables and notations that will be employed in the 
following.

Experimentally, GPDs are most simply accessed through the measurement of the
exclusive leptoproduction of a photon (Deep Virtual Compton Scattering -DVCS-)
and, possibly, of a meson (Deep Virtual Meson Production). In this short write-up,
we want to concentrate on the recent proton DVCS data that have emerged from two pionneering
experiments from the Halls A and B of Jefferson Laboratory and which provide
to this day the most extensive and precise data set for the DVCS process that has ever been 
available. The results of these experiments have recently been published~\cite{franck} or 
are under refereeing process~\cite{fx}. The goal of these conference
proceedings is humbly to attempt a first ``global" understanding of both these data sets.

\section{The JLab data}

\subsection{Hall A experiment E-00-110}

Hall A experiment E-00-110 has measured for the first time ever the cross section 
of the DVCS process in the valence region ($W\approx$ 2~GeV), the region of 
interest to access quark GPDs. The $ep\to ep\gamma$ reaction was
identified by detecting the scattered electron with the high resolution
arm spectrometer of the JLab Hall A and the final state photon in a PbF$_2$ crystals
calorimeter. The DVCS process was then identified by cutting on the
missing mass of the proton. Contamination by $ep\to e\gamma N\pi$ or
$ep\to ep\pi^0$ events could be estimated and subtracted, in particular by using 
a sample of fully exclusive events where the missing proton was actually detected 
in a dedicated array of scintillators.

\begin{figure}[htb]
\epsfxsize=4.cm
\epsfysize=7.cm
\epsffile{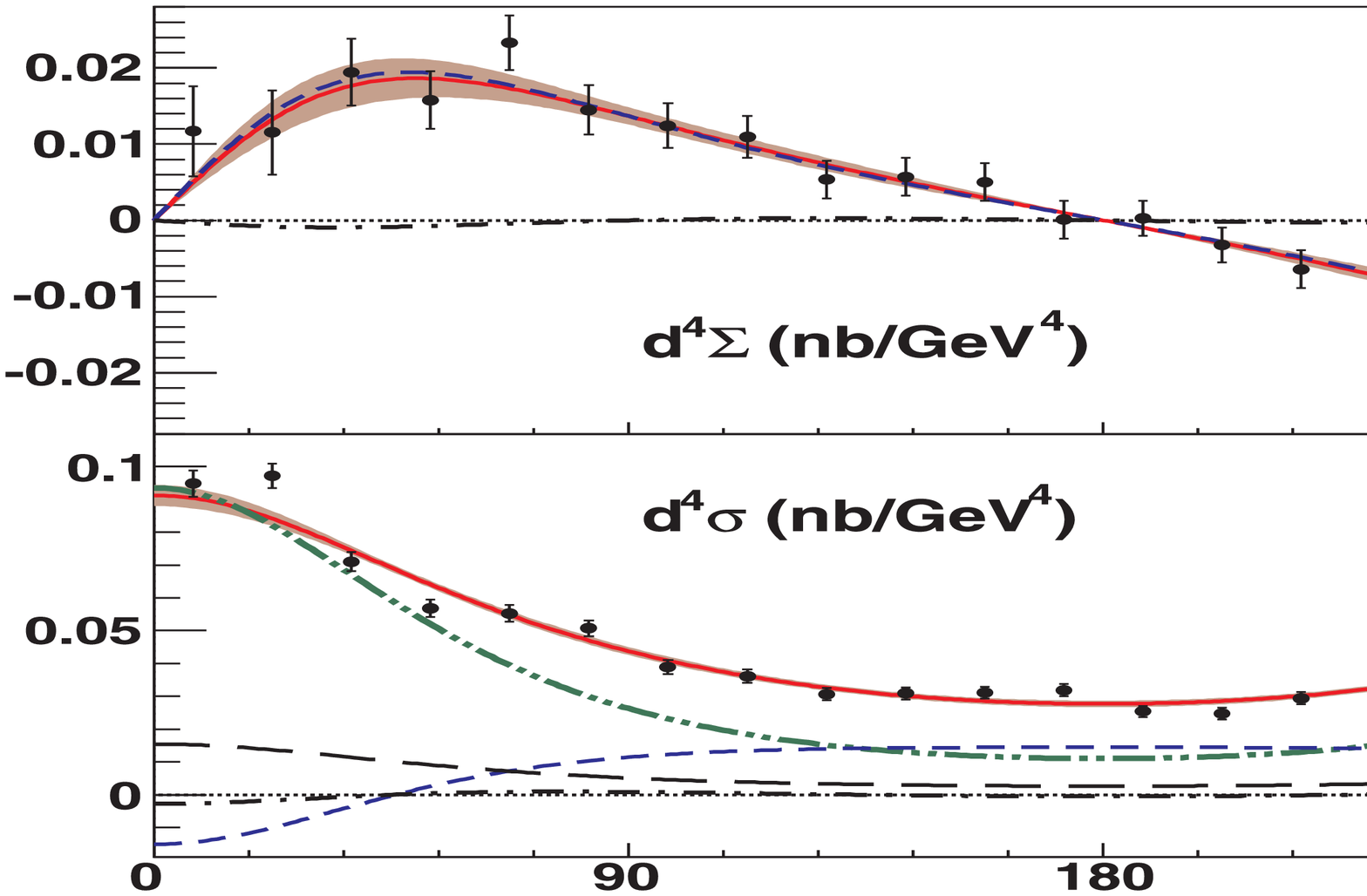}
\caption{The figure on the top shows the difference of (beam) polarized
cross sections for DVCS on the proton, as a function of the $\Phi$ angle, measured 
by the JLab Hall A collaboration~\cite{franck}. The average kinematics is
$<x_B>$=0.36, $<Q^2>$=2.3~GeV$^2$ and $<-t>$=0.28~GeV$^2$. The figure on 
the bottom shows the total (i.e unpolarized) cross section as a function of 
$\Phi$. The red curves show a fit to the data. The BH contribution is
represented by the dot-dot-dashed green curve. 
The difference between the data and the BH is attributed to the DVCS
whose twist-3 contribution is estimated by the dot-dashed curve (i.e., it is very small).}
\label{fig:halla}
\end{figure}

The 4-fold (polarized and unpolarized) differential cross sections $d\sigma\over dx_BdQ^2dtd\Phi$ 
(i.e. without any integration over an independent variable) have been extracted for
3 bins in $Q^2$ and are shown 
on Fig.~\ref{fig:halla} for $<Q^2>$=2.3~GeV$^2$, the highest $Q^2$ reached in the experiment.

The particular shape in $\Phi$ of the unpolarized cross section (lower panel of
Fig.~\ref{fig:halla}) is typical of the Bethe-Heitler (BH) process where the final 
state photon is radiated by the beam or the scattered electron. It is a process
which leads to the same final state than the DVCS process and which 
therefore interferes with it. The dot-dot-dashed green curve on the lower part
of Fig.~\ref{fig:halla} shows its precise shape and contribution. It can be seen that it
dominates most of the cross sections and only around $\Phi$= 180$^o$, there is 
a large discrepancy (a factor $\approx$ 2) between the BH and the data which could 
a priori be attributed to the DVCS process itself and therefore could contain sensitivity 
to GPDs. As it is an unpolarized cross section, this sensitivity is through the
square of the DVCS amplitude and both its imaginary and real parts are
expected to contribute.

The difference of polarized cross sections allows a different sensitivity to GPDs~:
on the one hand, like in general most of polarization observables, it is
sensitive to the imaginary part of the process, therefore to the sole imaginary part
of the DVCS process since the BH amplitude is real~; and on the other hand, since it arises
from an interference, it is sensitive in a linear fashion to the DVCS amplitude.

\subsection{Hall B experiment E-00-113}

\begin{figure}[h!]
\includegraphics[width=7.cm]{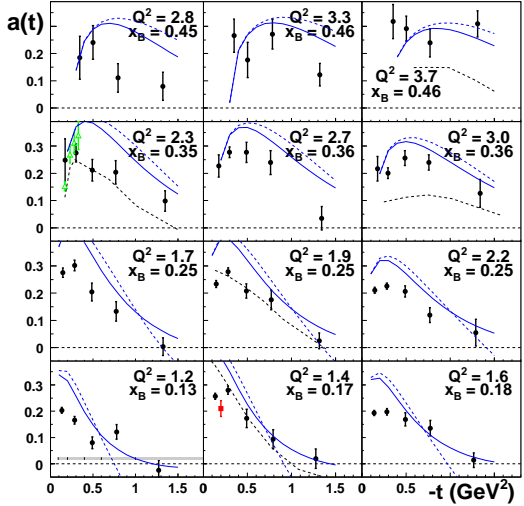}
\caption{Black circles~: beam spin asymmetry at $\Phi$ = 90$^o$ as a function of $t$ for different
($x_B$,$Q^2$) bins, as measured by the JLab CLAS collaboration~\cite{fx}. Green triangles
are the results extracted from the Hall A cross sections measurements~\cite{franck}. The 
red square is an earlier result from the CLAS collaboration~\cite{earlyclas}. The 
different curves are discussed in the text.}
\label{fig:bsa} 
\end{figure}

Experiment E-00-113 used the JLab Hall B CLAS spectrometer
to measure this same process $ep\to ep\gamma$. Here, the three final state
particles were detected, using, in particular, for the final state photon,
a new dedicated PbWO$_4$ crystals calorimeter equipped with Avalanche
Photo-Diodes. This did not fully prevent a residual contamination by
$ep\to ep\pi^0$ events but this latter background could be estimated 
($\approx 10\%$ in average) and subtracted by Monte-Carlo techniques.
Due to the large acceptance of the CLAS spectrometer, the largest-ever
phase space for DVCS in the valence region has been explored by this experiment. 
Fig.~\ref{fig:bsa} shows the beam spin asymmetries (BSAs) which have extracted
and which have been recently submitted to publication~\cite{fx}.

On this figure, the blue solid curves are the result of a DVCS GPD calculation using 
twist-2 Double Distributions for the $H$ GPD ($\tilde{H}$ and $\tilde{E}$ being neglected), 
based on the so-called VGG model~\cite{vgg1,vgg2}. The blue dashed curves are the result of 
the corresponding twist-3 Double Distributions calculation. As it will be
discussed more precisely in the next section, these calculations tend in general
to overestimate the data at low $t$. The dashed black curve is the result of a Regge 
calculation~\cite{jml} for the DVCS process, i.e. not based on GPDs, which, although
interesting in itself as providing a potential additionnal contribution to
the DVCS process, we will not discuss here.

We note that the BSAs which have been extracted from the
Halls A and B are relatively compatible and consistent, where the data overlap
($<x_B>$= 0.35, $<Q^2>$=2.3 GeV$^2$). The Hall B unpolarized cross sections and 
differences of polarized cross sections are still under analysis and should hopefully be 
expected before the end of this year. It will be crucial that similar agreement is reached
for these normalized cross sections. Let us recall that these pioneering experiments are 
definitely challenging, measuring 4-fold differential cross-sections at the level of a few 
picobarns.

\section{Data interpretation}

In this section, let us concentrate our discussion on the single $<x_B>$$\approx$ 0.36, $<Q^2>$=
2.3 GeV$^2$, $<-t>$= 0.28 GeV$^2$ bin, which is common to the Halls A and B experiments and
for which we have 3 observables available to compare with~: unpolarized cross
sections, differences of polarized cross sections and beam spin asymmetries (i.e. the ratio
of the two former ones).

In principle, two processes contribute to the $ep\to ep\gamma$ reaction in the kinematic
regime we discuss here~: the BH and the DVCS which should be added at the amplitude level.
The BH is very well known depending only on the nucleon form factors which are rather
well controlled at the rather low $t$ value discussed here. Therefore the lever arm
lies only in the DVCS process, both in its real and imaginary parts.
Being at fixed $x_B$, $Q^2$ and $t$ and, furthermore, the $\Phi$
dependence of the 3 observables being imposed on general grounds by the BH and 
leading-twist handbag DVCS amplitudes, 
the only freedom is therefore adding or subtracting a ``constant" (i.e. independent of $x_B$, $Q^2$, $t$ 
and $\Phi$) to either the imaginary or the real part of the DVCS amplitude.

\begin{figure}[h!]
\includegraphics[width=9.cm]{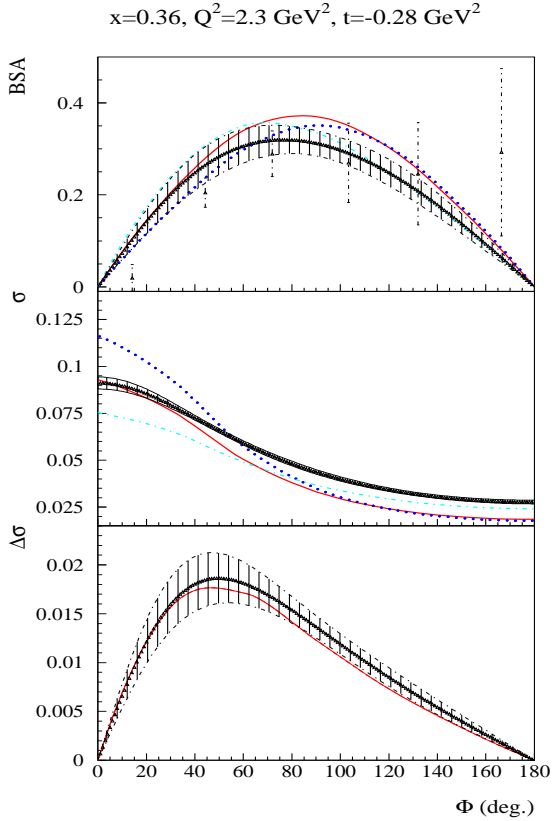}
\vspace{1.5cm}
\caption{Beam spin asymmetry -BSA- (upper panel), unpolarized cross section -$\sigma$- (middle panel)
and difference of polarized cross sections -$\Delta\sigma$- (lower panel) as a function of $\Phi$ 
for $<x_B>$= 0.36, $<Q^2>$=2.3 GeV$^2$ and $<-t>$= 0.28 GeV$^2$. The solid black line
is the result of a fit to the Hall A data and the band around it, the associated error.
The experimental points with the largest error bars in the upper panel are the Hall B data.
Note that the Hall B kinematics is slightly different than the Hall A's~: 
$<x_B>$= 0.34, $<Q^2>$=2.3 GeV$^2$ and $<-t>$= 0.30 GeV$^2$. The theoretical curves are discussed in
the text.}
\label{fig:theo} 
\end{figure}

As already mentionned, from Fig.~\ref{fig:bsa}, it seems that the standard Double 
Distributions VGG calculation systematically overestimates the experimental BSAs.
Two obvious remedies to this are then~: diminishing the imaginary part of the DVCS amplitude
or increasing the real part of the DVCS amplitude. Fig.~\ref{fig:theo} shows the
(simultaneous) effect of increasing the real part of the DVCS amplitude 
on the BSA, the unpolarized cross section and the difference of polarized cross section.

Let us first concentrate on the upper part of Fig.~\ref{fig:theo} which 
shows the BSA for the above mentionned $<x_B>$= 0.36, $<Q^2>$=
2.3 GeV$^2$, $<-t>$= 0.28 GeV$^2$ bin as a function of $\Phi$.
The red solid curve is basically the VGG twist-2 calculation of Fig.~\ref{fig:bsa}
which confirms here that (both the Halls A and B) experimental data are overestimated.
The dotted (dark) blue line shows the effect of ``boosting" the real part of the DVCS amplitude.
A typical motivation for this would be the presence of a strong D-term~\cite{weiss} 
(in this calculation, we have taken, as a simple example, the Weiss-Polyakov D-term 
multiplied 
by $F_1$(t) for its $t$ dependence and renormalized by a factor 2). We see that, as anticipated,
this allows the BSA to decrease and be, at least at $\Phi=90^o$, in better agreement with the data.
This figure shows by the way that the analysis of the BSA cannot simply be reduced to 
the value of the BSA at $\Phi=90^o$ as the whole $\Phi$ dependence shape, which is
not a simple sin$\Phi$, can show some sensitivity and information to the DVCS amplitude.

Now, getting an agreement simply for the BSA is not so much a challenge as it is 
a ratio and several contributions can compensate or annihilate each other. What is the effect on the 
other observables ? The middle panel of Fig.~\ref{fig:theo} shows that
neither the ``standard" VGG twist-2 calculation (red curve) nor the ``real part-modified"
are able to describe the unpolarized cross section. If the ``standard" VGG twist-2 
calculation matches the data at $\Phi=0^o$, it misses by a factor $\approx$ 2 the
data at $\Phi$= 180$^o$. The ``real part-modified" calculation even removes the
agreement at $\Phi$= 0$^o$. 

On Fig.~\ref{fig:theo}, the solid (light) blue curve shows the result of changing the sign of 
the additional real part term to the DVCS amplitude (i.e. taking an opposite sign to the one 
advocated in Ref.~\cite{weiss}). The BSA at $\Phi=90^o$ also decreases and
brings the theoretical calculation in relative agreement with the data. More
precisely, the general $\Phi$ dependence of the BSA seems to show a better agreement with the Hall A 
data and somewhat less with the Hall B data. Turning to the unpolarized cross section, it gives a nice
agreement in the region $\Phi$= 180$^o$ but underestimates strongly the data at 
$\Phi$= 0$^o$. In the framework of this short study, whatever the value and the sign of 
this real term, the conclusion should be clear~: it is  impossible to reproduce the 
$\Phi$ dependence and normalisation of the unpolarized cross section.

We finally note that the difference of polarized cross sections (lower panel
of Fig.~\ref{fig:theo}) is insensitive to the presence or not of a D-term (be it strong
or not, positive or negative)~: all three calculations just discussed give 
the same result (thin solid red line) as it is an observable sensitive only to the imaginary 
part of the DVCS amplitude. The agreement for this latter observable is very good and 
this tends to give support to the option of decreasing the real part of the DVCS amplitude in 
order to get an agreement for the BSA, if it were not for the strong disagreement for the 
unpolarized cross section that we just observed.

The other option that can be envisaged to reconcile the theory and the data for the
BSA is, as mentionned at the beginning of this section, to decrease the imaginary 
part of the DVCS amplitude. We do not discuss precisely this solution here but it has
recently been pursued in details by M. Vanderhaeghen and M. Polyakov~\cite{marcmax}.
In the framework of the ``dual" model, it can be shown that the imaginary part of
the DVCS amplitude can be decreased in a natural way. These authors have shown that
the agreement with the BSA is very satisfying (like it is by increasing
the real part as we just saw), but that, again, no agreement with the $\Phi$
dependence of the unpolarized cross section can be found (the disagreements are
at the same level of what we show in the middle panel of Fig.~\ref{fig:theo})
and, furthermore, the difference of polarized cross sections is now in severe
disagreement.

\section{Conclusion}

For the first time in the field of GPDs, we are confronted with strong experimental constraints 
due to the large data sets that have recently poured and been released by the 
JLab Hall A and CLAS collaborations (with much more to 
come soon !). For the first time, theory might be faced with a strong challenge. This short study 
has shown that, at this date, with a few models that have been developped so far (Double 
Distributions, D-term, dual model), it is basically impossible to reproduce, even for a 
single $<x_B>$, $<Q^2>$ and $<-t>$ bin, simultaneously the $\Phi$ dependence of the three 
``most basic" observables  which are the BSA, the unpolarized cross section and the difference 
of polarized cross section for DVCS on the proton. Given that the lever arm in any model is 
very limited (changing the real or the imaginary part of the DVCS amplitude at a fixed 
$<x_B>$, $<Q^2>$ and $<-t>$, the $\Phi$ dependence being imposed by the
general forms of the leading-twist handbag DVCS and BH amplitudes), it seems that this 
conclusion might be rather model independent.
Further data (Hall B cross sections data in particular are eagerly waited) are clearly 
needed in order to resolve this issue. We finally stress that this short study
has been carried in the approximation of neglecting the $E$, $\tilde{H}$ and $\tilde{E}$ GPDs
(justified in the framework of the VGG model). If the strength of these latter GPDs
turned out to be much more important than the VGG predictions, some conclusions
might possibly have to be reconsidered.


\begin{footnotesize}

\end{footnotesize}

\end{document}